\documentclass[english,reprint,showpacs,superscriptaddress,preprintnumbers,nofootinbib,amsmath,amssymb,prl,floatfix]{revtex4-1}
\usepackage{graphicx}
\usepackage{nima,color}
\newcommand{\pa}{\pr}

\begin{document}
\title{Crises and Physical Phases of a Bipartite Market Model}
\author{ Nima Dehmamy}
\affiliation{CCNR, Northeastern University, Boston, MA 02115}
\author{Sergey Buldyrev}
\affiliation{Department of Physics, Yeshiva University, New York, New York 10033, USA}
\affiliation{Center for Polymer studies Boston University,
Boston, MA 02215 }
\author{Shlomo Havlin}
\affiliation{Bar-Ilan University, 52900 Ramat-Gan, Israel}
\author{H. Eugene Stanley}
\affiliation{Center for Polymer studies Boston University,
Boston, MA 02215 }
\author{Irena Vodenska}
\affiliation{Administrative
Sciences Department, Metropolitan College, Boston University, Boston, MA
02215 USA,}
\affiliation{Center for Polymer studies Boston University,
Boston, MA 02215 }

\begin{abstract}
    We analyze the linear response of a market network to shocks based on
    %We show that
    the bipartite market model we introduced in \cite{dehmamy2014classical},
    which we claimed to be able to identify the time-line of the 2009-2011 Eurozone crisis correctly. We show that this model
    has three distinct phases that can broadly be categorized as ``stable'' and ``unstable''.
    Based on the interpretation of our behavioral parameters, %While
    the stable phase describes periods where investors and traders have confidence in the market (e.g. predict that the market rebounds from a loss). We show that the unstable phase happens when there is a lack of confidence and seems to describe ``boom-bust'' periods in which changes in prices are exponential.
    We analytically derive these phases and where the phase transition happens using a mean field approximation of the model.
    %We derive the phases of a bipartite model for markets, which we introduced in an earlier paper \cite{dehmamy2014classical} in which we were able to identify the correct time-line of the 2009-2011 Eurozone crisis. Here we show that the phases of this model can be broadly categorized as ``stable''  and ``unstable''.
    %We argue that both may occur in real market situations, though the unstable phase can only last for short periods of time. The unstable phase may indicate both periods of rapid growth and rapid decline (i.e. crisis).
    We show that the condition for stability is $\alpha \beta <1$ with $\alpha$ being the inverse of the ``price elasticity'' and $\beta$ the ``income elasticity of demand'', which measures how rash the investors make decisions.
    We also show that in the mean-field limit this model reduces to a Langevin model \cite{bouchaud1998langevin} for price returns.
    %We also validate the existence of these phases and the phase transition between them by calculating the full phase diagram of the Eurozone crisis data as initial conditions.
    Thus we  provide analytical support for the power of the model in classifying the stability of markets where it's applicable.
\end{abstract}
\maketitle

\section{Inroduction}
Economical and financial systems are attractive to physicists because of their highly dynamic and complex structure which is generally much more challenging as a complex system than systems exhibiting a high degree of symmetry studied in traditional physics. The goal of this paper is to derive analytical results regarding the physics and phases of a model we developed for quantifying the linear response in a bipartite, dynamical network. While our discussion will exclusively concern a largely simplified financial network and its mean-field behavior, a similar methodology based on an ``effective theory'' approach could be employed in analyzing the response of many other dynamical bipartite networks. Especially networks pertaining to resources and agents using resources, such as power grids and some special food webs, might benefit from a similar type of modeling.

The global financial crisis that followed which started around 2007 made it clear to us how limited our understanding of the dynamics of financial markets are. In recent years, economists seem to have acknowledgement the fact that we cannot predict the real world dynamics with idealized assumptions and we need to account for complex relations of different players in an economical system. Scholars have started consider the network of such interactions and how they may affect business cycles, cause cascades, or contagion. \cite{Gale2,acemoglu2012network}.
Our goal here is to make a first attempt at quantitatively modeling the continuous dynamics of a market system in the simplest form and to lowest order.
In an earlier paper we introduced a bipartite network model for investment markets in which investors traded assets in a fashion similar to common stock markets \cite{dehmamy2014classical}. The assumption of the market being a bipartite network is, of course, a major simplification. In reality, there undoubtedly exist many strong financial interactions among investors and even among assets. In our paper \cite{dehmamy2014classical} the major investors were banks or other major financial institutions. It should be noted that interbank lending network is a very complex and important, multiplex network and much research has been dedicated to it (see \cite{furfine2003interbank,upper2011,bargigli2016interbank, farboodi2014intermediation} and references therein).

Our simplified bipartite model, which has ``assets'' as one layer and ``investors'' on the other side, ignoring all intra-layer connections, was constructed in order to assess the first order response dynamics of a market to any change. In this model neither money nor number of shares is conserved, as investors are assumed to trade with entities both inside and outside the network -- though such conservation laws can be imposed if needed. The weighted connections indicate the amount of investments. The details of the model and the phenomenological derivation for it are explained in the \cite{dehmamy2014classical}. The goal of this paper is to derive analytical results about the stability conditions for such market dynamics. %Here we shall concern ourselves with the rich mathematical and physical properties of the model, which

There are many different models for financial networks \cite{Haldane,allen2000financial,Kok,kok2, kok3} each assessing a different aspects of connectivity which use familiar assumptions about dynamics in of prices based on existing economic models.
Our model, on the other hand, is at its core what we call an ``effective theory'' in physics. It attempts to model the dynamics of a market by making minimal assumptions about details and assuming system-wide parameters. It only uses the microscopic behavior of agents as input. The demonstration of the genericity of our model is the subject of another forthcoming paper \cite{lagrangian}.

Perhaps the closest model to ours in the context bipartite model of investment networks was introduced by Caccioli et al. \cite{caccioli2014stability}. As will become clear in the derivations below, our model in fact also reduces to a Langevin equation's model for price-dynamics in a stock market introduced by Bouchaud \cite{bouchaud1998langevin}. There the authors argue that the price dynamics near the onset of a crisis is simlar to a damped harmonic oscillator plus non-linear terms that become relevant after the phase transition to an unstable phase. Our model has such non-linear naturally from the start and no extra assumption is needed. In another forthcoming paper we will discuss the application of our model in stock markets in detail.

Our model relies on 2 behavioral parameters, one for supply/demand and one for investor decisions. Over  time-scales where the behavioral parameters are not changing much, we are arguing that the large scale dynamics of the market can modeled and simulated. It is in this regime that our model may provide cucial insight into stability of a market and serve as a tool for regulators.

\section{Model and Notation}
We will refer the model as
the Group-Impulse Portfolio-Sharing Investment (GIPSI) model henceforth. In essence the GIPSI model summarizes how investors and their brokers may behave on average in a market, if they are aware of the market news.

\begin{figure}
\centerline{\Large\color{blue}Investors}
\centerline{\includegraphics[width=.5\columnwidth%1.7in
, trim=0 0 0 1.5cm, clip]{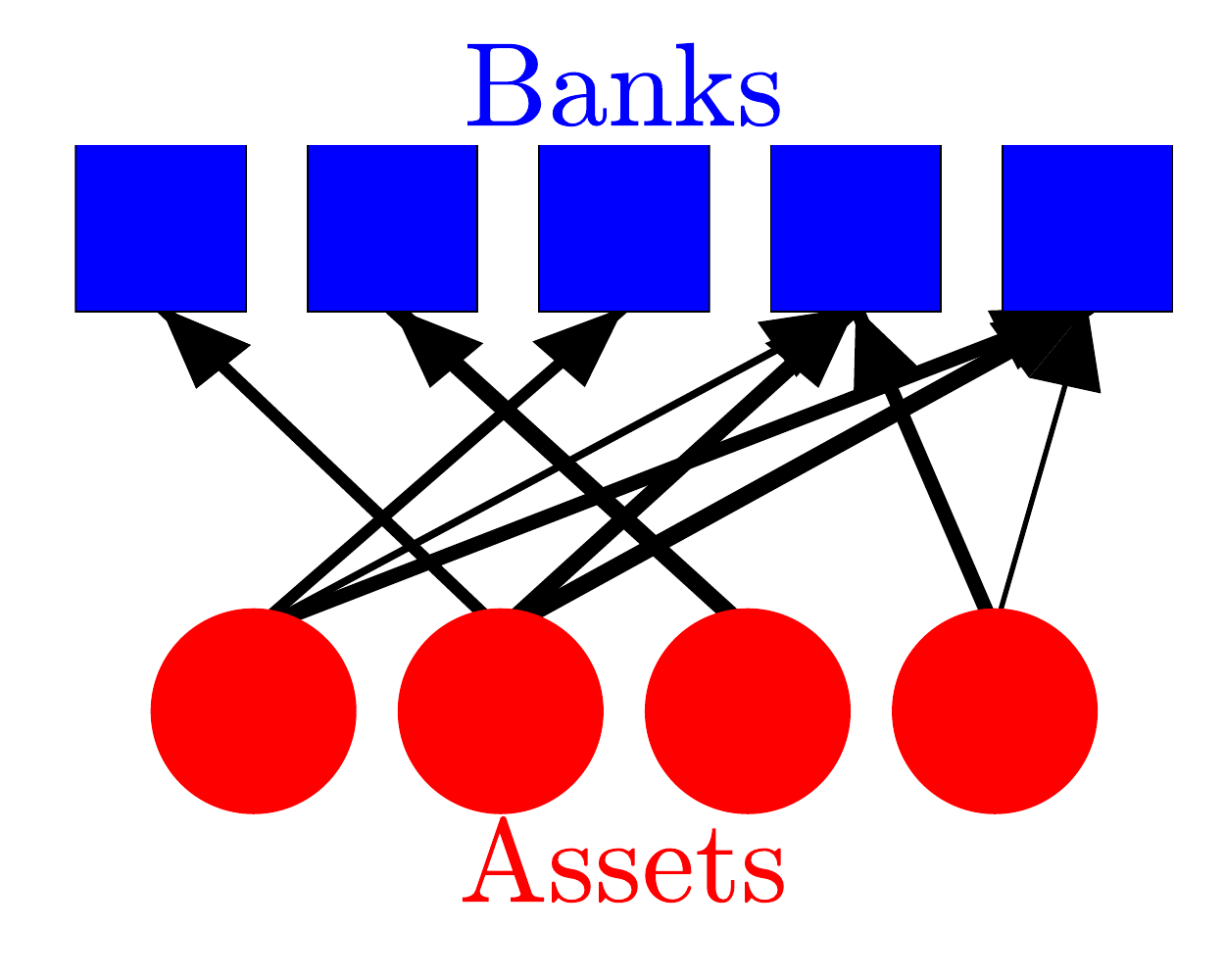}
}
\caption{A sketch of the network of investors vs assets. It is a
  directed, weighted bipartite graph (thicknesses represent
  weights of investment). The wighted adjacency matrix has  entries $A_{i\mu}$, the number of shares of asset $\mu$ held by investor $i$. \label{fig:graph}
} % caption

\end{figure}

We will approximate the investiment market as a bipartite network as shown in
Fig.~\ref{fig:graph}.  On one side we have the ``Assets,'' which are what is traded, and on the other we have the ``Investors'' that
own the assets. The ``assets'' are labeled using
Greek indices $\mu,\nu...$.  To each asset $\mu$ we assign a ``price,''
$p_\mu(t)$ at time $t$. The ``investors'' are labeled using Roman
indices $i,j...$. Each investor has an ``equity'' $E_i(t)$, a time $t$, i.e. their net worth. Each investor has a portfolio, meaning
differing amounts of holdings in each of the asset types. The amount of asset $\mu$ that investor $i$ holds is denoted by $A_{i\mu}(t)$, which is
essentially an entry of the weighted adjacency matrix $A$ of the
bipartite network.

We assume that there is a group-impulse in the decision-making, the so-called ``herding effect'' \cite{devenow1996rational,bikhchandani2000herd} where all agents will have the same level of panic or calmness in a situation. They will adjust their portfolio $A_{i\mu}$ by looking at the gains or losses $\delta E_i$ they incurred recently. So one of the equations is a generalization of a simple portfolio adjustment protocol
\[\delta \log A_{i\mu} \sim \beta \delta \log E_i \]
We argue that, if the assumption
Their level of panic is reflected in the so-called ``Income elasticity of demand'', which we will denote as $\beta$.
This assumption has been critisized by some scholars as being unrealistic based on other empirical evidence about ``fire sales'' \cite{shleifer2011fire}, claiming that the leverage (i.e. ratio of investments to equity) . Our justification for this assumption is based on a mean field assumption. We believe that most of these institutions will be leveraged to close to the maximum amout allowed by regulations, thus having similar leverage. Aside from that, if the assumption of herding holds for response to a rapid change, using the same factor $\beta$ makes sense for at least the mean field behavior of the system.

The supply and demand equation for the prices $p_\mu$ has a similar structure
\[\delta \log p_\mu \sim \alpha \delta \log A_{i\mu} \]
where $1/\alpha$ would be what is usually called ``price elasticity'' in economics. There is also a third equation which is related to how trading, i.e. $\delta (A\cdot p)_i$  changes the equity of an investor, which turns out to be \cite{dehmamy2014classical} $\delta E_i = (A\cdot \delta p)_i$. A crucial point for having a more realistic model for such systems is the fact that there is a response time associated with each of these equations.
describing how each of the variables $E_i(t), A_{i\mu}(t)$, and
$p_\mu(t)$ evolve over time. A key feature of our model is that the
weights of links $A_{i\mu}$ are time-dependent, and this introduces
dynamics into our network.

For brevity, we define $\ro_t\equiv {d\over dt}$. The equations of the GIPSI model may be written as
\begin{align}
\pa{\tau_B\ro_t^2 +\ro_t }A_{i\mu}(t)&=\beta {\ro_t E_i(t)\over E_i(t)} A_{i\mu}(t)\label{eq:ddA}\\
\pa{\tau_A\ro_t^2 +\ro_t } p_\mu(t)&=\alpha {\ro_t A_\mu(t)\over A_\mu (t)}p_\mu (t)\label{eq:ddp} \\
\ro_t E_i(t)&= \sum_\mu A_{i\mu}(t) \ro_t p_\mu(t)+f_i(t). \label{eq:ddE}
\end{align}
where $f_i=dS_i/dt$ has the meaning of external force.
where $\tau_B$ is the time-scale in which investors respond to a change in their net worth,
and $\tau_A$ is the time-scale of market's response.

The importance of this model as a lowest order approximation of linear response in a market system is the subject of a separate paper in preparation \cite{lagrangian,dehmamy2016graduate}. But just to motivate the use of this model, we only note that the lowest order effective Lagrangian model (in the spirit of Landau-Ginzburg models) for the response of a market defined by the dynamical variables $A_{i\mu}, p_\mu$ and $E_i$ plus dissipation yields equations with the structure of \eqref{eq:ddA}-\eqref{eq:ddE}.

\begin{table}
\caption{Notation\label{tab:definitions}}
%\end{table}
%\begin{table}
\centering
\begin{tabular}{@{\vrule height 10.5pt depth4pt  width0pt}cc}
\hline
symbol & denotes \\
\hline
$A_{i\mu}(t)$ & Holdings of investor $i$ in asset $\mu$ at time $t$\\
$p_\mu (t)$ & Normalized price of asset $\mu$ at time $t$ ($p_\mu(0)=1$)\\
$E_i(t)$ & Equity of bank $i$ at time $t$. \\
$\alpha$ & Inverse price elasticity\\%``Inverse market depth'' factor of price to a sale.\\
$\beta$ & Income elasticity of demand (rashness)\\
\hline
\end{tabular}
\end{table}

%\outNim{120714 %%%%%%%%%%%%%%%%%%%%%%%%%%%%%%%%%
\section{Confidence in the market in terms of based on market response}
The simple strategies devised above are expected to describe the linear response of the market in the mean-field approximation. The sign of $\beta$ may be considered an indicator for confidence in the market: positive $\beta$ means in response to a loss (i.e. $\delta E_i <0$) the investor sells assets (i.e. $\delta A_{i\mu} <0$). This can be interpreted as the investor fearing more losses and therefore reducing their holdings. By the same token negative $\beta$ may signal confidence in the market as it indicates a willingness to buy more shares when the investor has lost money.

Note that the GIPSI model Eq. \eqref{eq:ddA}--\eqref{eq:ddE} are response euation, i.e. they yield no dynamics of there is no change in the variables $E,A,p$.
To assess the behavaior of this system we will introduce an initial shock to the system.
We do this by assuming a delta function change in equity $f_i(t) = c \delta(t)$ for some investor $i$, meaning that at $t=0$ agent $i$ either gained or lost money from outside the market and it's decision to trade in the market triggers the response of the market \footnote{It turns out that the qualitative behavior of the final state after a shock to the system is only weakly sensitive to the value of the response times $\tau_A,\tau_B$. The response times only need to be nonzero. As we will show analytically below, the phase of the system is determined by the two behavioral couplings $\alpha$ and $\beta$. The plots shown below are for $\tau_A = \tau_B = 1$.}.

For the empirical analysis we will use the same Eurozone crisi data we used in the original paper \cite{dehmamy2014classical}
which consists of sovereign bonds of five European countries, namely
Greece Italy, Ireland, Portugal and Spain  (GIIPS), as assets and the network of major banks and financial institutions who purchased and traded these bonds.
Many smaller players and the European Central Bank (ECB) that also participated in the trades are not in this data and their existence justifies the non-conservation of the total amount of holdings in the network.
They are also partly responsible for dissipation in the dynamics. The data is given in the appendix of \cite{dehmamy2014classical}.

\begin{figure*}
\centering
\includegraphics[width=.95\columnwidth]{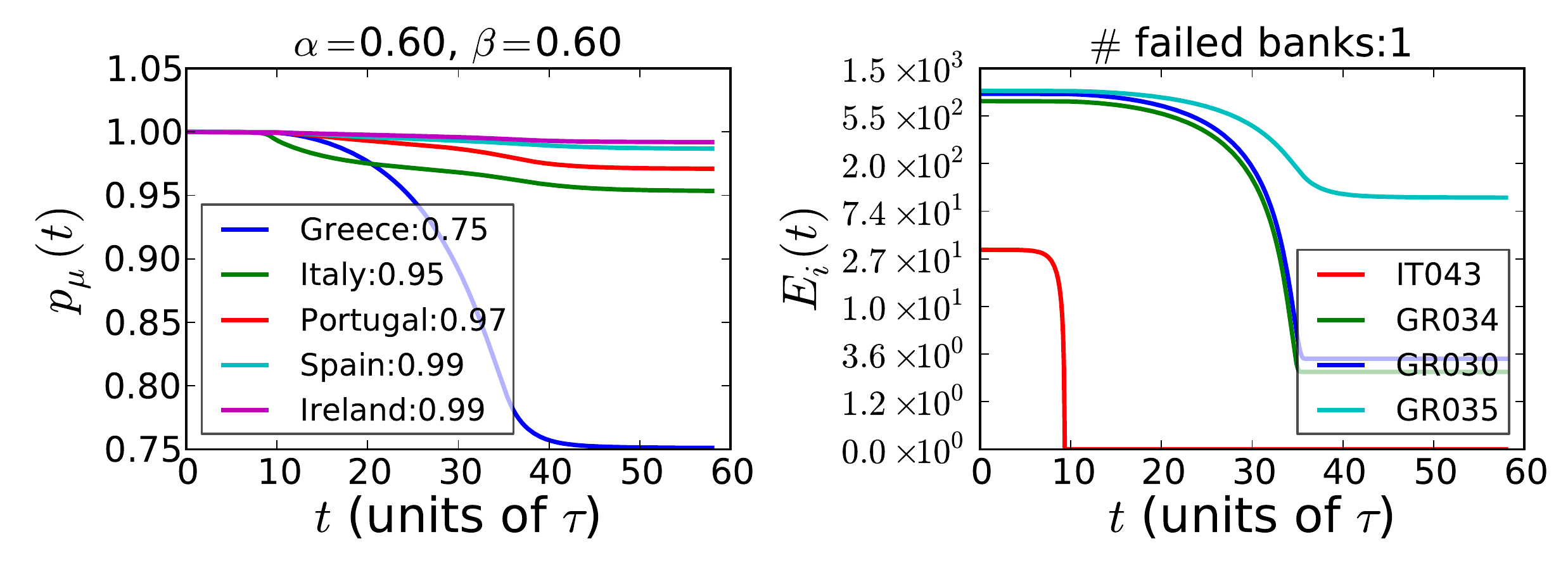}
\includegraphics[width=.95\columnwidth]{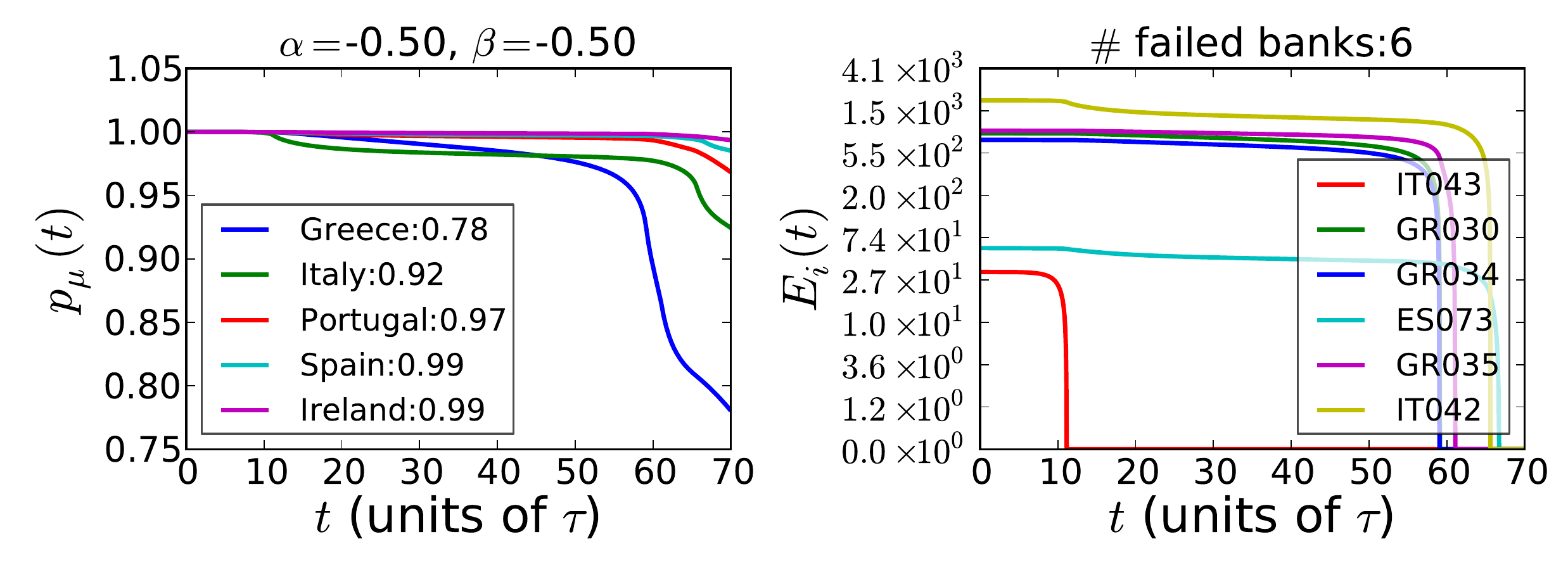}
\includegraphics[width=.95\columnwidth]{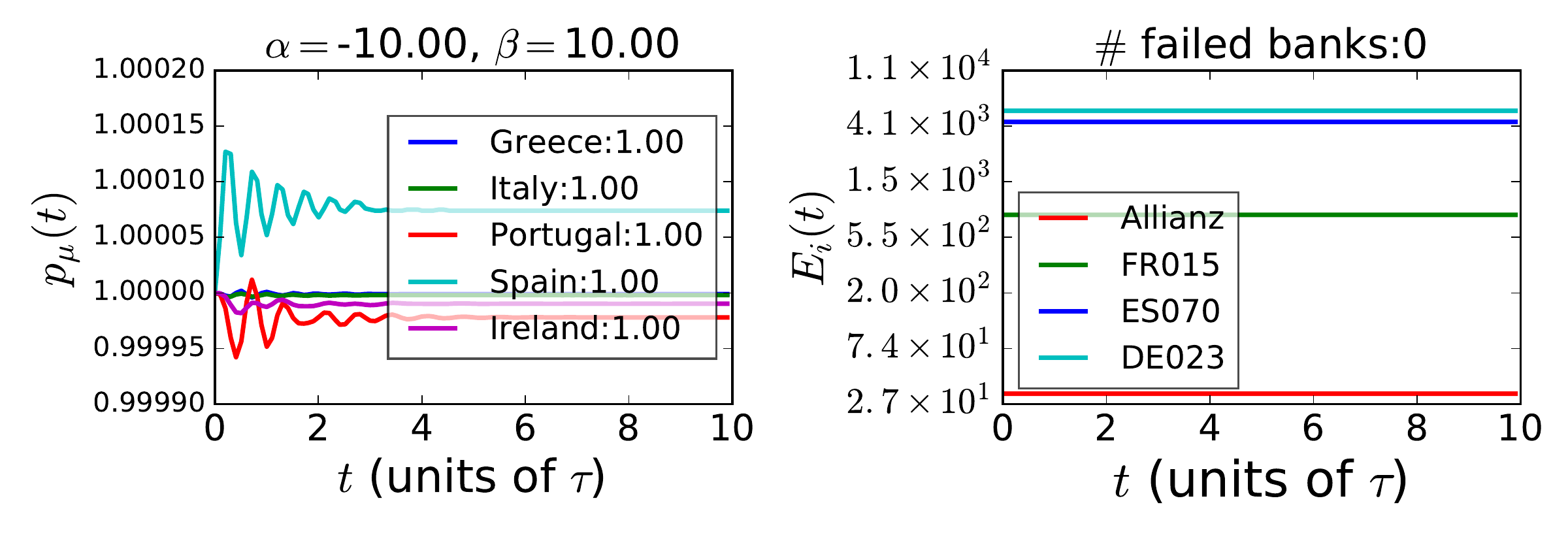}
%pEEBAanb100.pdf}
\includegraphics[width=.95\columnwidth]{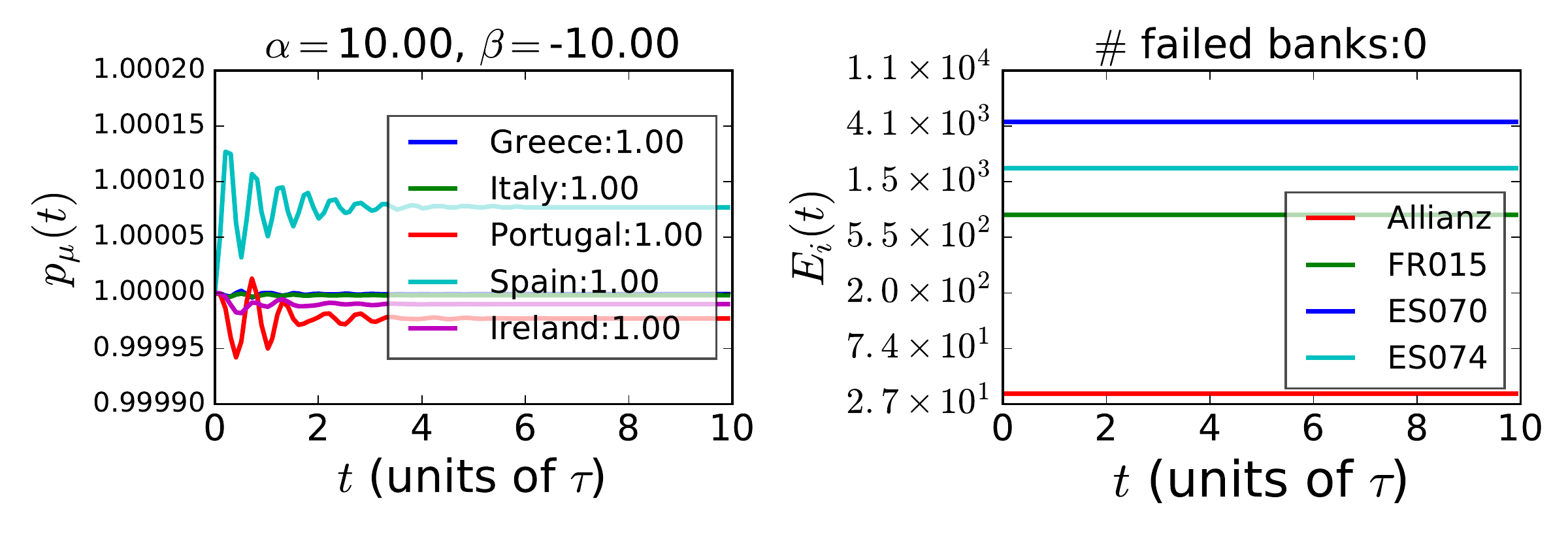}
%pEEBAabn100.pdf}
\caption{Contrarian regimes: top, both $\alpha,\beta<0$. Here many banks
  fail, even for relatively small $\alpha, \beta$. The losses are
  devastating. Our model suggests that such a regime should be
  avoided. The bottom two plots show the two points $\alpha=\pm \beta$,
  $\beta=\pm 10$. The two results are almost identical. They also show
  that no appreciable amount of profit or loss is generated in these
  regimes, thus making them rather unfavorable for investors most of the
  time, but because of their safeness could be a contingency plan
  (buyout of bad assets by central banks is one such contrarian
  behavior).
\label{fig:4}}
\end{figure*}

Fig. \ref{fig:4} shows 4 examples of choices for $\alpha$ and $\beta$ and the response of the prices and equities of banks to a shock to a random bank in the system
\footnote{It turns out that, similar to a system in the thermodynamic limit, the details of which bank is shocked through $f_i$ doesn't matter in the final state of the system\cite{dehmamy2014classical}.}.
As we will show below, all these choices fall in the ``stable'' regime, though the losses incurred from a negative shock
(i.e. $f_i< 0$) are much larger in magnitude when $\alpha$ and $\beta$ have the same sign than when they have opposite signs.
The upper left plot in Fig. \ref{fig:4} is for $\alpha$ and $\beta>0$. This is what one normally expects from this system: $\beta>0$
means if a bank incurs a loss, they try to make up for it by making
money from selling assets; $\alpha>0$ means if there is selling
pressure (more supply than demand) the prices will go down. There are,
however, cases where the opposite happens.
Negative values for $\alpha$ and $\beta$ are possible in the real world and are known as ``contrarian'' behavior.
For example, a contrarian investor is someone who invest more in asset $\mu$ when they actually lose money on asset $\mu$.
This happens a lot when there is confidence in the market and investors believe the price will rebound.
%``Contrarian'' agents in a market are those who, for example, buy more assets when they incur losses, hoping to recover some of their losses by reducing the average cost of investment.
The market may also sometimes behave in a contrarian fashion, when there is an anticipation of good news that
overcomes the selling pressure, or when other investors outside our Eurozone GIIPS
network (such as smaller investors or the ECB) are
actually exerting a buying pressure.

As is seen in Fig. \ref{fig:4} this results in price slightly oscillating around its original value. This trust mechanism plays an important role in the stability of the market.
It requires that one side (either investor or the prices) behaves in a contrarian fashion and the other side behaves normally.

\section{Phases and Confidence}
\begin{figure*}%[ht]
\centering
\includegraphics[width=.5\columnwidth]{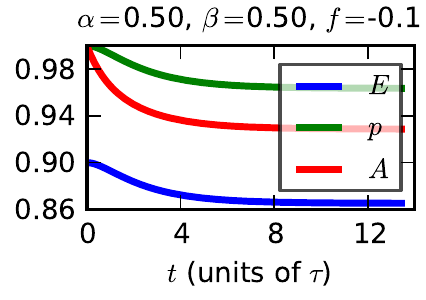}\includegraphics[width=.5\columnwidth]{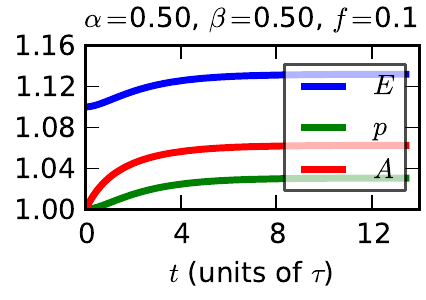}
\includegraphics[width=.5\columnwidth]{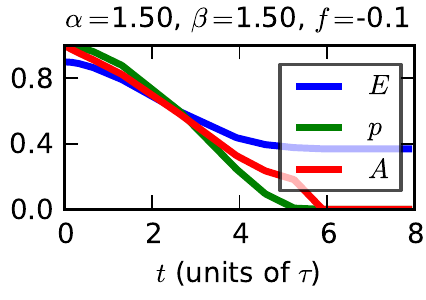}\includegraphics[width=.5\columnwidth]{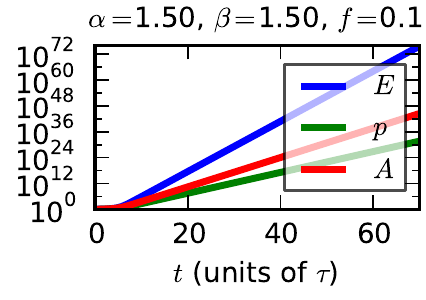}
\outNim{\includegraphics[width=1.7in]{1ddamped.pdf}\includegraphics[width=1.7in]{1d-damped.pdf}
\includegraphics[width=1.7in]{1ddiv.pdf}\includegraphics[width=1.7in]{1d-div.pdf}
}
\caption{Numerical solutions to the differential equations in a 1 bank
  vs 1 asset system. The upper plots show a ``stable'' regime, where
  after the shock none of the variables decays to zero or blows up, but
  rather asymptotes to a new set of values. The lower plots are in the
  ``unstable'' regime where positive or negative shocks either result in
  collapse or blowing up or collapsing of some variables.
\label{fig:1d}}
\end{figure*}

As we see the behavior of prices $p_\mu$ and equities $E_i$ in the case where the product $\gamma \equiv \alpha\beta$ is positive and negative  differ qualitatively. When $\gamma >0$ negative shocks ($\ro_t E_i(0) <0$) cause a noticeable drop in prices and some equities\footnote{All of $E,A,p$ are positive quantities and we do not allow them to go negative.}, whereas when $\gamma < 0$ there is only oscillation and the final values of $p_\mu$ and $E_i$ are very close to their original values.
\begin{figure}
    \centering
    \includegraphics[width = \columnwidth]{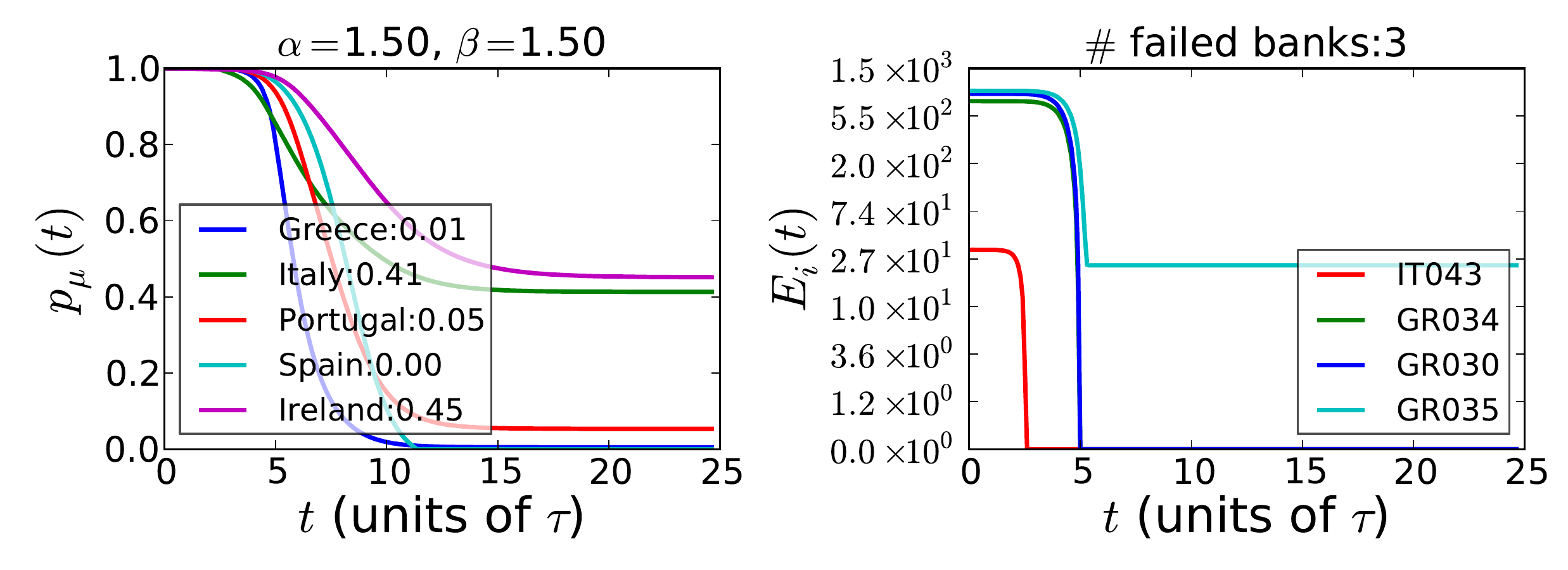}
    \caption{When $\gamma = \alpha\beta >1$, although the shape of the price and equity plot after a negative shock look similar to when $1>\gamma>0$, the behavior of the system is qualitatively different. Many asset prices plummets down to or close to zero. The system survives only because investors (banks) whose equity goes to zero stop existing and won't propagate the shock further. }
    \label{fig:unstable}
\end{figure}
There is, however, another qualitatively different regime which is not shown in Fig. \ref{fig:4}. This happens when $\gamma >1$ and is shown in Fig. \ref{fig:unstable}. In fact, this qualitative difference becomes much clearer when we show the effect of a positive shock $\ro_t E_i(0)>0$ to the system in Fig. \ref{fig:1d}. There we see that when $\gamma<1$ after a positive shock eventually stops growing and reaches a new equilibrium, whereas when $\gamma>1$ all three variables $E,A,p$ grow exponentially, indefinitely. In other word an economic ``bubble'' forms. So $\gamma>1$ if it lasts for a long time either forms a bubble or results in a crash.
% Stable ones are those where the initial shock is dampened quickly and the
% system goes to a new equilibrium, without any of the variables $E, A, p$
% either collapsing exponentially to zero or blowing up
% exponentially.
% Such behaviors in response to sudden rise or sudden fall
% in $E$ in a 1 investor vs 1 asset system is shown in figure \ref{fig:1d}.
%

This leads to the following conclusions:
\begin{enumerate}
\item When $\gamma=\alpha\beta <0$
  no investor goes bankrupt, but also the amount of money lost or generated during the trading is negligible.
  This makes these regimes (where either the investors or the market is contrarian,
  but not both) good for preventing failures, but they are very
  undesirable for profit making.
\item When $1>\gamma >0$ the system does not show oscillation but eventually settles into new equilibrium states not far from the original situation. Negative shocks may cause bankruptcy of some investors, depending on how their assets $(A\cdot p)_i$ compares to their equity $E_i$ initially.

\item When $\gamma >1$, negative shocks cause exponential drops in assets and equities. If $\gamma<1$ persists there will be a crisis. Positive shocks may form bubbles and periods of exponential growth.

\end{enumerate}

\subsection{The full phase diagram}

\begin{figure}
\centerline{
\includegraphics[trim = 1cm 0 20mm 0,  clip,
  width=.5\columnwidth]{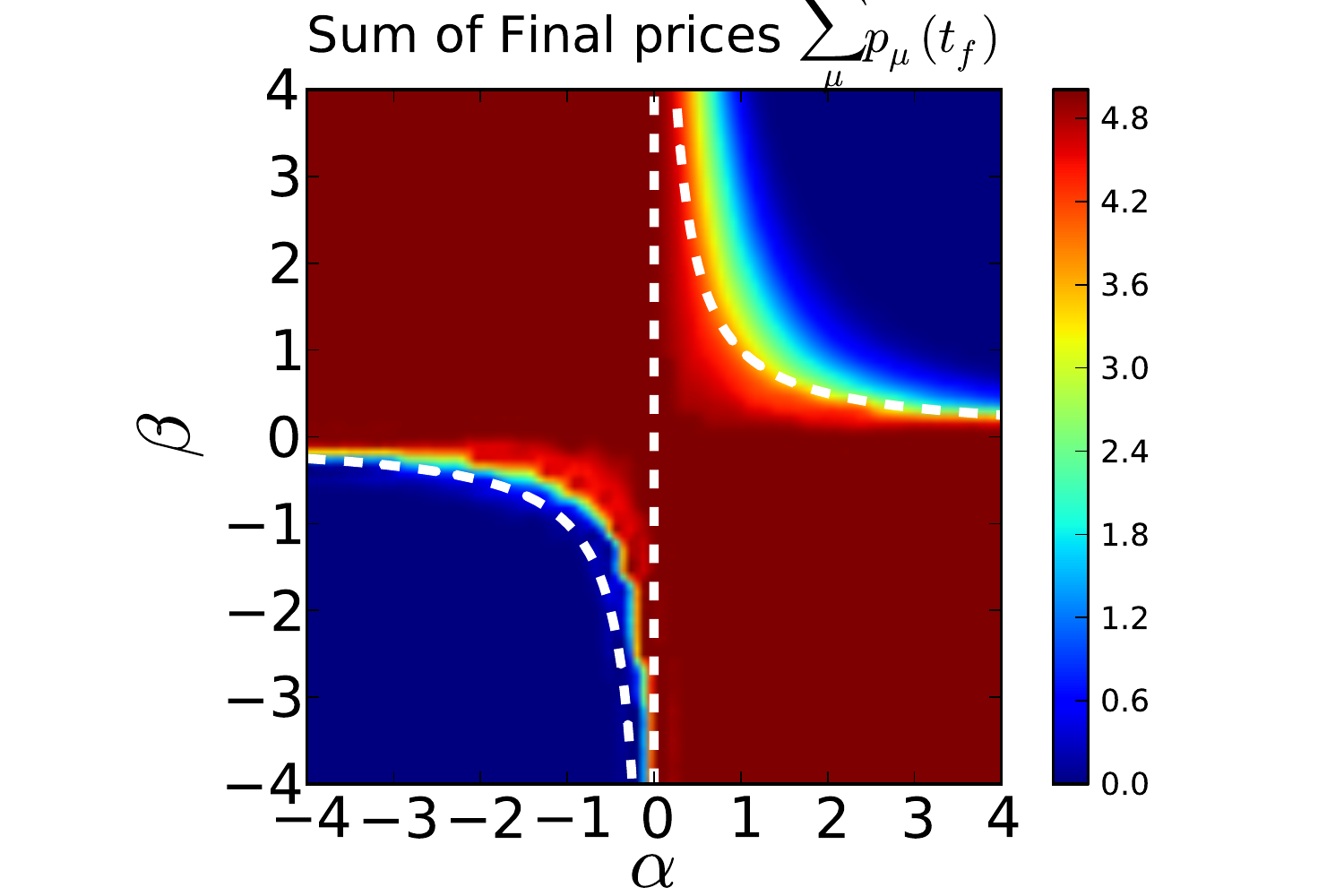}\includegraphics[trim = 1cm 0
  20mm 0, clip,width=.5\columnwidth]{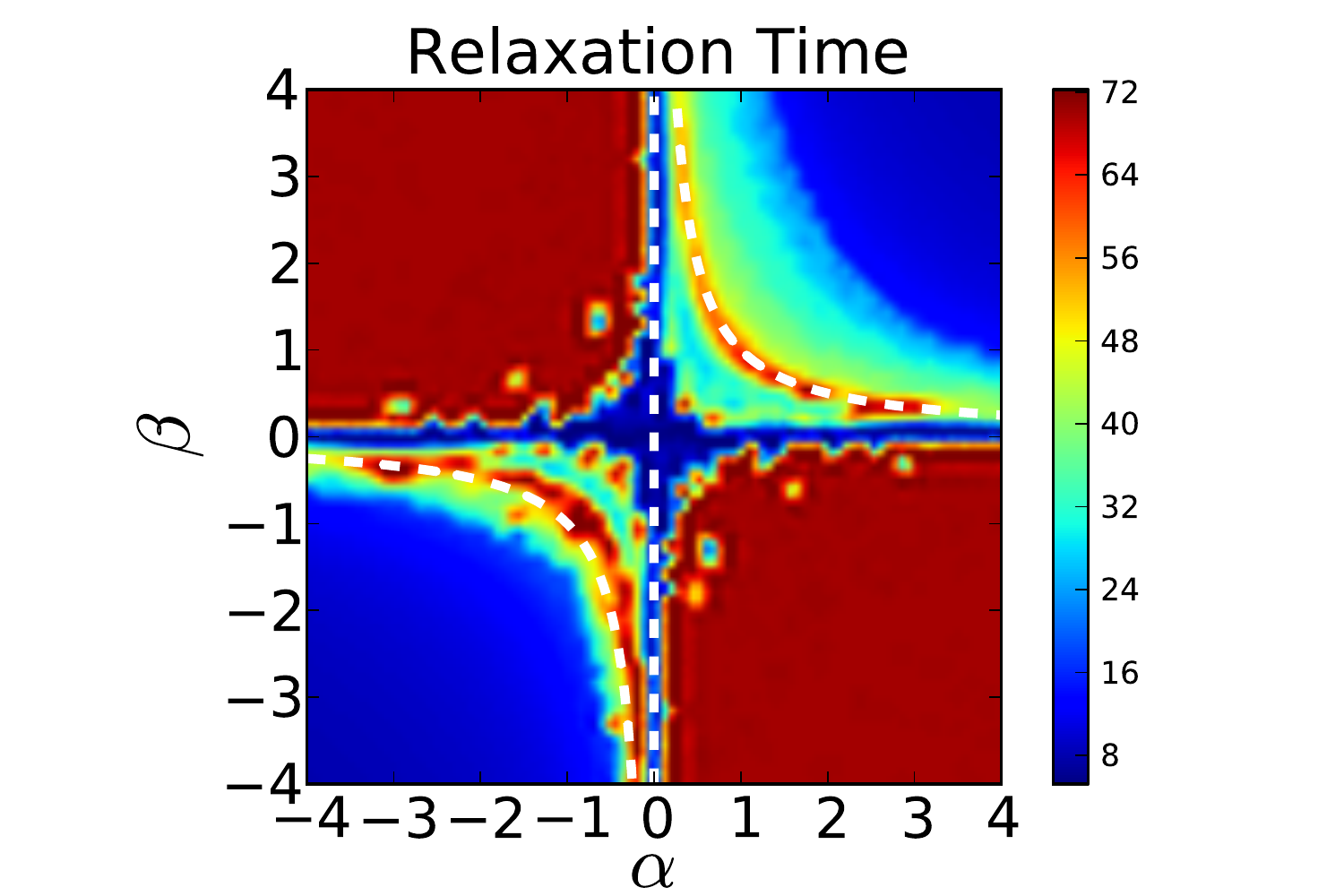}}
\caption{{\bf Left}: Phase diagram of the GIIPS sovereign debt data, using the
  sum of the final price ratios as the order parameter.  We can see a
  clear change in the phase diagram from the red phase, where the
  average final price is high to the blue phase, where it drops to
  zero. The drop to the blue phase is more sudden in the $\alpha<0,
  \beta<0$ quadrant than the first quadrant. {\bf Right}: The time it takes
  for the system to reach the new equilibrium phase. This relaxation
  time significantly increases around the transition region, which
  supports the idea that a phase transition (apparently second order)
  could be happening in the first and third quadrants.  The dashed white
  line shows the curve $\gamma=\alpha \beta =1$. It fits the red curves
  of long relaxation time very well. This may suggest that $\gamma=1$ is
  a critical value which separates two phases of the system.
 \label{fig:phase}}
\end{figure}

Fig. \ref{fig:phase} shows an example of the average final prices and
relaxation time for the system for various values of $\alpha$ and
$\beta$. It seems the system has two prominent phases: One in which a
new equilibrium is reached without a significant depreciation in all of
the GIIPS holdings (upper left and lower right quadrants), and one where
all GIIPS holdings become worthless (above dashed line in the upper
right quadrant and all of lower left quadrant).
%In the third quadrant the transition is much more abrupt than in the first quadrant.
In both the first and the third
quadrants in the transition region the relaxation time becomes very
large, which means that the forces driving the dynamics become very
weak. Both the smoothness and the relaxation time growth seem to be
signalling the existence of a second order phase transition.  The phase transition %in the first quadrant
seems to be described well by:
\[\gamma=\alpha \beta =1.\]

\outNim{082016
But this result is not exact and below we derive a more precise form for
this equation, which is:
\begin{equation}
\gamma = 1+ f_0,
\end{equation}
where $f_0$ is the magnitude of the initial shock $f_i(t)= f_0\delta(t)$
for a fixed $i$ that's being shocked.
082016}

Now we do a systematic numerical analysis of different phases of this phenomenological model. We identify to phases and what appears to be a second order phase transition between them.
We then modify the equations \eqref{eq:ddA}--\eqref{eq:ddE} and analytically derive the condition for the phase transition. The GIIPS data is very heterogeneous, in the sense that for each of these European countries there are generally less than a handful of  investors who hold more than half of the total sovereign debt of that country.
If we make a mean field assumption and break the network apart, assuming there is one major investor for each asset $\mu$, we can treat that as a 1 investor 1 asset system and try to derive the phases in that case.
As we show below, the 1 by 1 system, despite not having the richness of the networked system in terms of the details of the final state, nevertheless exhibits the same phases. In addition we are able to analytically derive the phase transitions.

% From examining the simulations more closely and from numerical analysis
% of the differential equations \eqref{eq:ddp},\eqref{eq:ddA} and
% \eqref{eq:ddE} in networks of few nodes, presented below, we see that as
% expected the equations have either stable or unstable solutions.

%
%\outNim{
%120614
%

\section{Analytical derivation of the mean-field phase space}
In the mean-field approach described above, we are dealing with a 1 investor by 1 asset system with a single $E,A$ and $p$. We can combine Eqs. \eqref{eq:ddA}--\eqref{eq:ddE} in a 1 by 1 system by taking another $\ro_t$ derivative from \eqref{eq:ddp}. This way we can eliminate most occurrences of $E$ and $A$ an find an equation for $p$ (see Appendix \ref{ap:eqs1}) which has some non-linear terms in it.
% \begin{align}
% \left[\tau_A\tau_B \ro_t^2+(\tau_A+\tau_B
%     )\ro_t+\pa{1-\gamma {Ap\over E}} \right]\ro_t p = O\pa{(\ro_t
%     p)^2} \label{eq:d3p}
% \end{align}

Defining the ``return'' $u \equiv \ro_t p$ as the fundamental
variable, the nonlinearities are roughly of type $u^2 + a \ro_t u^2$.
In short, the equations are
\begin{align}
\left[\tau \ro_t^2+\ro_t+\omega^2 \right]u  &={O\pa{u^2,\ro_t u^2}\over p}\cr
{1\over \tau}= {1\over \tau_B}+{1\over \tau_A}, %\cr{}
\quad &\omega^2= {1-\gamma{Ap\over E}\over \tau_B+\tau_A}.%\cr{}
%p(t)&= \int^t u(t') dt'
\label{eq:d3p}
\end{align}
As we show in Appendix \ref{ap:eqs1} none of the nonlinearities on the r.h.s. can be large in the stable phase where $\gamma < 1$. We also show there that $Ap/E$ becomes its $t=0$ value plus other nonlinear terms that cannot be large when $\gamma < 1$ and
\({Ap\over E} \approx 1+f_0 +O(u^2). \)
For a small shock $f_0 = -\eps$ we may safely use $Ap/E = 1$.
Thus for a time-scale where $\gamma <1 $ and is not changing much we are essentially dealing with a damped harmonic oscillator. Notice that equation \eqref{eq:d3p} is almost identical to what Bouchaud proposes in \cite{bouchaud1998langevin} to explain the 1987 crash.

Although $\omega^2$ depends on $A,p$ and $E$, we can use an approximate
exponential ansatz $u\sim u_0 \exp[\lambda t]$. The
solutions to $\lambda$ are:
\[\lambda_\pm={-1\pm \sqrt{1- 4 \tau \omega^2}\over 2\tau}\]
Thus we get three regimes:
\begin{enumerate}
\item When $\omega^2>{1\over 4\tau}$
  there will be oscillatory solutions. This happens when $\gamma < {-(\tau_A-\tau_B)^2\over \tau_A\tau_B}.$ For $\tau_A = \tau_B$ this is just the $\gamma < 0$ condition we observed for oscillations in our simulations. %For example when $\gamma{Ap\over E}<-1$, which only happens for negative $\gamma$ we have such oscillatory solutions. This is consistent with the simulations which showed the oscillatory behavior was in the $\alpha\beta<0$ quadrants. For the stability, however we care about the real solutions.
\item When ${1\over 4\tau}>\omega^2
  >0 $, i.e.  ${-(\tau_A-\tau_B)^2\over \tau_A\tau_B } < \gamma < 1$  we have decaying solutions but both $\lambda_\pm <0$. Therefore the changes won't be large and eventually the sytems settles in new equilibrium.
\item When $\omega^2<0$, i.e.
 $\gamma>1$ we will have two real solutions for with opposite signs. The presence of the positive root $\lambda_+>0$ signals an instability because this solution diverges.

\end{enumerate}
Thus we have proven the existence of the three phases we had observed earlier and dervied the transition conditions analytically. It is easy to prove that in the unstable phase both exponents $\lambda_\pm$ appear in the solution:
\outNim{082016
For
a delta function shock of magnitude $f$ at $t=0$ we found that:
\[E_0\to E_0(1+f)\]
Having initially scaled to $E_0=A_0=p_0=1$, the condition for existence of the positive root becomes:
\[t=0:\quad \gamma> {E\over Ap}= (1+f)\]
This dependence on the shock magnitude is normal, as a strong enough
kick can kick a particle out of a local minimum.  The shock can be
arbitrarily small and therefore the absolute condition for stability is
as we anticipated
\begin{equation}
\large \mbox{\bf unstable at:} \quad \gamma >1 \label{eq:unstable}
\end{equation}

Now the question is, which solution does the system pick when it is shocked.
082016}
The return $\ro_t p$ is
\[\ro_t p(t) = u(t)= u_+ e^{\lambda_+ t}+u_- e^{\lambda_- t} \]
Since at $t=0$ the initial conditions dictated $\ro_t p(0)=0$ we have
\(u_+=-u_-\)
and therefore both solutions appear with equal strength. It follows that
whenever one of the solutions ($u_-$ in our case) is positive the
solution diverges. When $f>0$ a bubble forms and grows exponentially and
when $f<0$, because our variables are non-negative, the price just
crashes to zero.  This proves that the sufficient condition for
stability is $\gamma<1$. Also note that the nonlinear terms are all
proportional to $\ro_t p$ and therefore at $t=0$
\[O\pa{u^2(0),\ro_t u^2(0)}=0\]
and so the solution is exact at $t=0$. The details of the derivations as well as the nonlinear terms are given in Appendix \ref{ap:eqs1}.

In Appendix \ref{ap:exp} we provide another, direct proof for the phase transition being at $\gamma=1$ using exponential ansatz for all three variables $X=E,p,A$
\[X\sim X_0 +X_1\exp[w_{X1} t]+X_2\exp[w_{X2} t].\]
We then show that near the phase transition the exponents in $E,A,p$  show that the near the phase exponents for all three need to be the same. Moreover, as can be seen in the right figure in Fig. \ref{fig:phase} near the phase transition the time-scale of the system's evolution is very long compared to other time-scales in the system and thus much longer than $\tau$ in \eqref{eq:d3p}, meaning
\[|w_X|\ll \tau^{-1} \]
This leads to a massive simplification of the equations and yields
\[\gamma = {1+f_0\over 1-\beta f_0} \]
as the phase transition point, which for infinitesimal shocks $f_0\to 0$ recovers $\gamma =1$.

\section{Conclusion}
We have shown that the GIPSI model \cite{dehmamy2014classical} for response dynamics of a bipartite investment market exhibits three phases: two stable phases, one of which is oscillatory the other logistic-like; one unstable phase with indefinite exponential decay or growth depending on the initial perturbation. The unstable phase can exist in real world, though it must be short-lived. It can describe ``boom-bust'' periods: crisis periods (as was shown in \cite{dehmamy2014classical}) or periods of rapid growth, so-called ``bubbles''. We also derive these phases analytically from a mean-field version of the model. The mean-field equations become almost identical to the Langevin equations proposed by Bouchaud \cite{bouchaud1998langevin} which could well describe behavior of markets near crashes. The non-linear terms required for the transition in Bouchaud's model exist naturally in the GIPSI model and nothing new needs to be added by hand.

\section{acknowledgments}

We thank the European Commission FET Open Project ``FOC'' 255987 and
``FOC-INCO'' 297149, NSF (Grant SES-1452061), ONR (Grant
N00014-09-1-0380, Grant N00014-12-1-0548),DTRA (Grant HDTRA-1-10-1-0014,
Grant HDTRA-1-09-1-0035), NSF (Grant CMMI 1125290), the European
MULTIPLEX and LINC projects for financial support.  We also thank
Stefano Battiston for useful discussions and providing us with part of
the data. The authors also wish to thank Matthias Randant and others for
helpful comments and discussions, and especially Fotios Siokis for
sharing important points about the data and the Eurozone crisis. S.V.B. thanks the Dr. Bernard W. Gamson Computational
Science Center at Yeshiva College for support.

\bibliographystyle{plain}
\bibliography{references}

%\end{document}

\newpage
~
\newpage
\appendix

\outNim{082016

A phase diagram
using $\ro_t E$ of the 1 by 1 system is shown in figure
\ref{fig:1dphase}.

\begin{figure}
\centering
\includegraphics[width=.4\columnwidth]{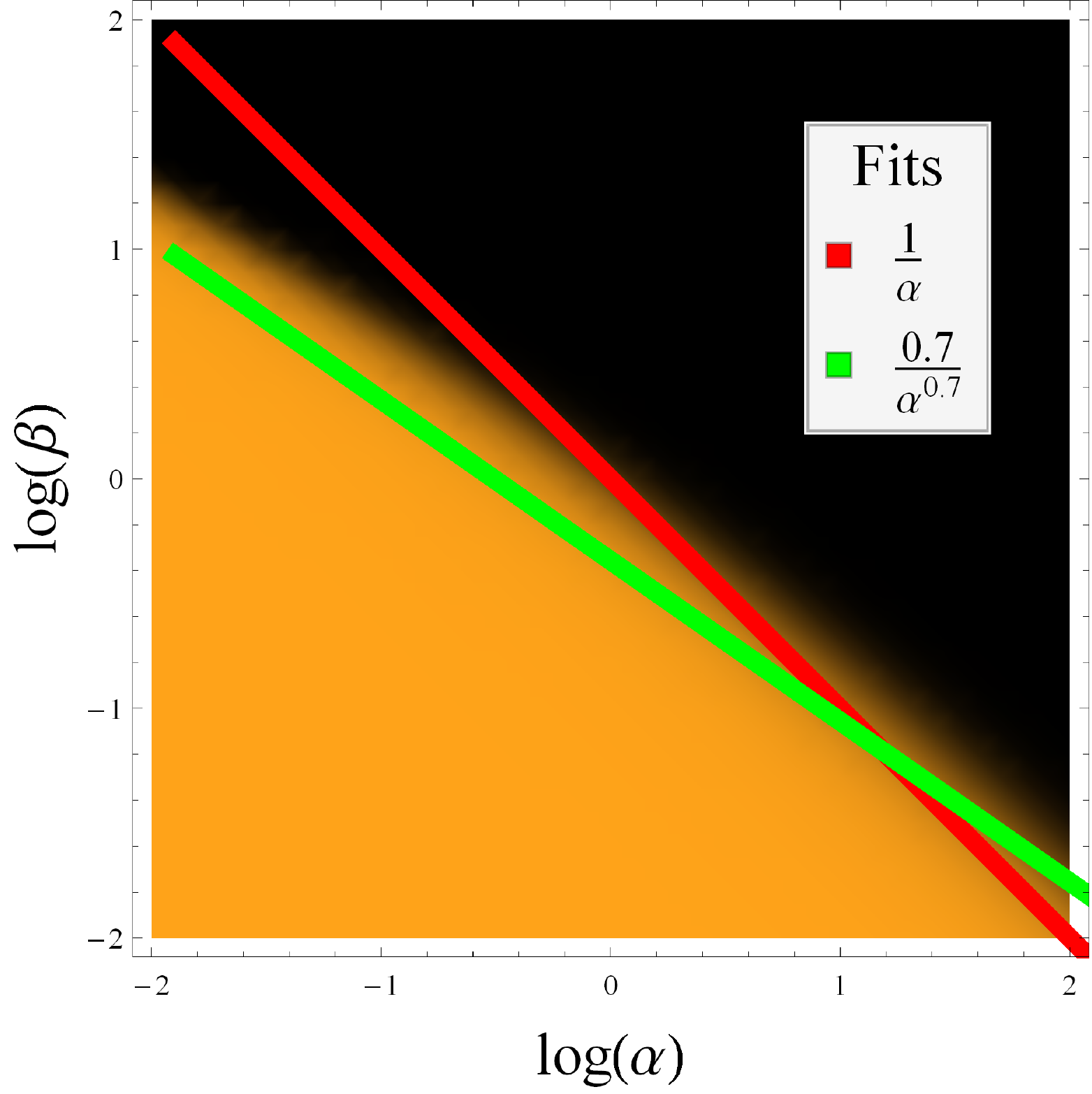}\includegraphics[width=.4\columnwidth]{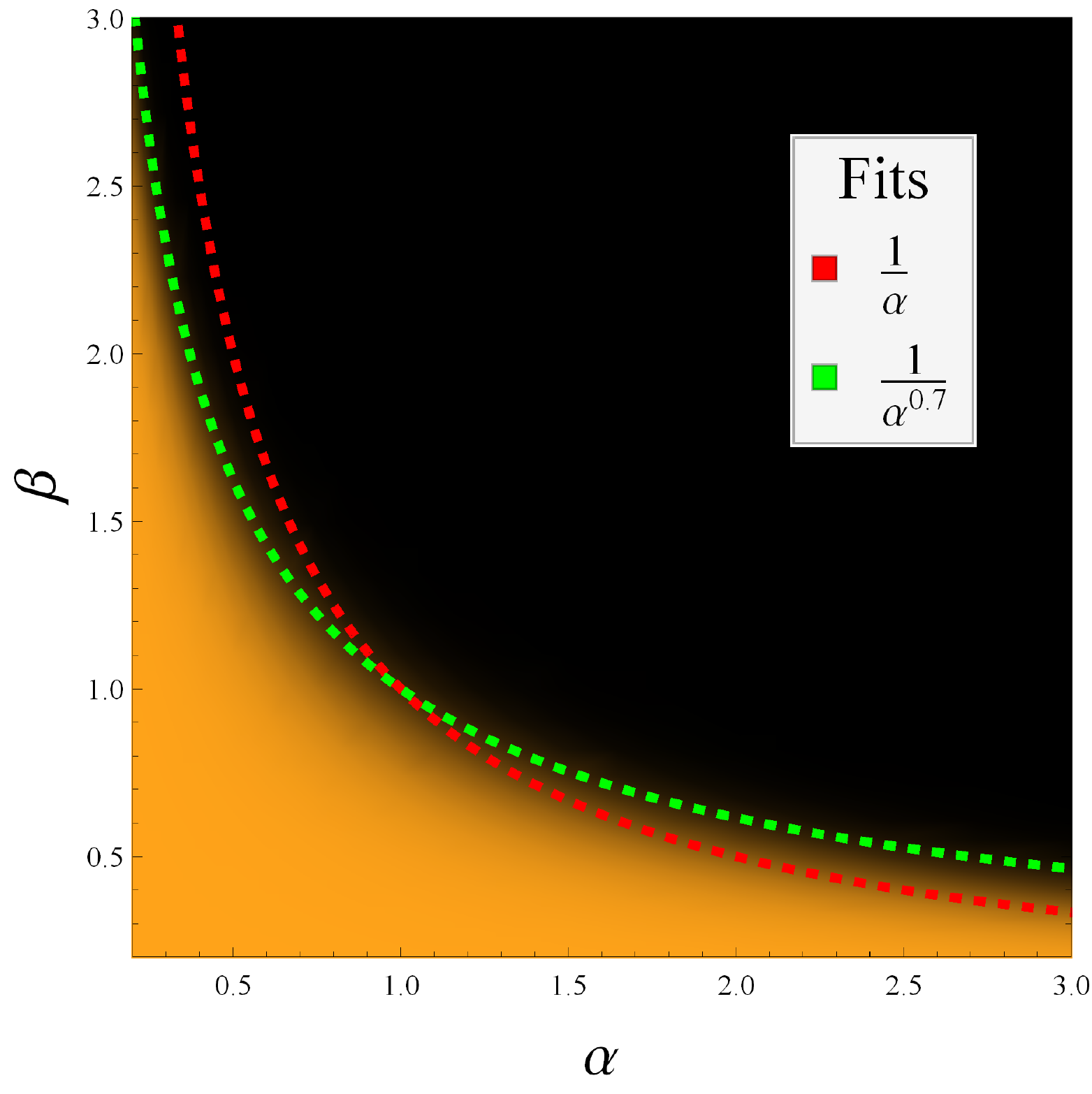}
\caption{Phase diagram of the 1 bank vs 1 asset system, responding to
  sudden rise in $E$. Black denotes regions where $\ro_t E=A \ro_t p$
  was very large at late times, and light orange where it was close to
  zero. This is only plotting the $\alpha,\beta>0$ quadrant (left is a
  log-log plot, right is the regular linear scale diagram). The overlay
  are two fit functions for the phase transition curve. While
  $\alpha\beta=1$ is not a very good fit for large $\beta$ and small
  $\alpha$, it fits fairly well for large $\alpha$'s and we analytically
  prove this below.
\label{fig:1dphase}}
\end{figure}

082016}

\section{Analytical results from the 1 Investor vs 1 Asset system\label{ap:eqs1}}

\outNim{082016

\begin{figure*}
\centering
\includegraphics[width=5.5in]{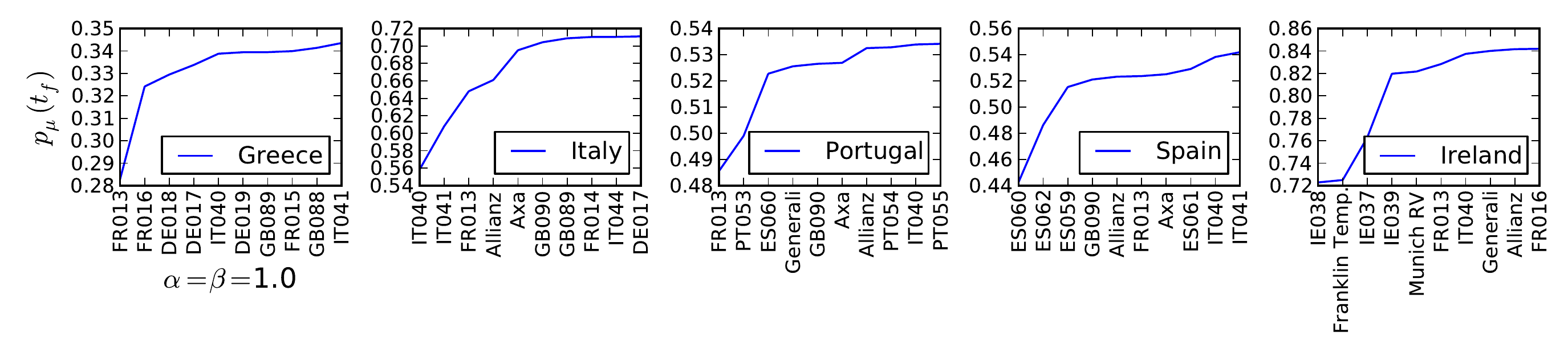}
\caption{Top 10 banks whose failure causes the most damage to the price
  of each country's sovereign bond. \label{fig:Rank10} }
\end{figure*}
Fig.\ref{fig:Rank10} shows the
top 10 banks whose failure at $\alpha=\beta =1$ causes the largest damage
to each of the 5 GIIPS assets.

082016}

Here we present the analytical solution to the 1 by 1 model and derive
the curve where the phase transition is happening in figure
\ref{fig:phase}. At any time $t$ the equations for a 1 by 1 system
become
\begin{align}
{(\ro_t+\tau_B\ro_t^2) A\over A}&=\beta {\ro_t E\over E}=\beta {A \ro_t p\over E}\cr
{(\ro_t+\tau_A\ro_t^2)p\over p}&=\alpha {\ro_t A\over A} \label{eq:1d1}%\cr
%\ro_t^2 E &= \ro_t A \ro_t p +A \ro_t^2 p
\end{align}
Below we will try to find the condition for a phase transition in the solutions to these equations.

%120614
%}

%\outNim{%%%%%%%%%%%%%%%%%%%%
%%%%%%%%%%%%%

We can try to eliminate $A$ and $E$.  We first need to find the
expression for $\ro_t^2 A/A$ first. Taking another derivative from the
second equation yields
\begin{align}
{(\ro^2_t+\tau_A\ro_t^3)p\over p}-{(\ro_t+\tau_A\ro_t^2)p\ro_t p\over p^2}&=\alpha {\ro_t^2 A\over A}-\alpha \pa{{\ro_t A\over A}}^2 %\cr
%\ro_t^2 E &= \ro_t A \ro_t p +A \ro_t^2 p
\end{align}
combining this with the first equation results in:
\begin{align}
&{(\ro_t+\tau_A\ro_t^2)p\over p}+\tau_B
  {(\ro^2_t+\tau_A\ro_t^3)p\over p} =\gamma {A \ro_t p\over E }
  +O\pa{(\ro_t p)^2}\cr &\left[\tau_A\tau_B \ro_t^2+(\tau_A+\tau_B
    )\ro_t+\pa{1-\gamma {Ap\over E}} \right]\ro_t p = O\pa{(\ro_t
    p)^2}
\end{align}
where the nonlinear term is again quadratic in $p$ (thus a generalized
form of the Fisher equation) and looks like
\begin{align}
O\pa{(\ro_t p)^2}=&\tau_B {(1+\tau_A\ro_t)\ro_t p\ro_t p\over p}\cr
&-\alpha \tau_B {\pa{(1+\tau_A\ro_t)\ro_t p}^2\over p}
\end{align}
Below we will also show that in the stable regime the non-linearity in the frequency, namely the $\gamma Ap/E$ term, is of the order $O(\ro_t A \ro_t p)$ and thus remains small if we show that at small times the behavior of $\ro_t p$ in the stable regime is oscillating around zero.

This time the dynamics is richer and we have a damped oscillator with a
driving force coupled to $p$ and nonlinearities of type $\sim (\ro_t
p)^2$. Taking the return $u \equiv \ro_t p$ as the fundamental
variable, the nonlinearities are roughly of type $u^2 + a \ro_t u^2$.
In short, the equations are
\begin{align}
\left[\tau \ro_t^2+\ro_t+\omega^2 \right]u  &=O\pa{u^2,\ro_t u^2}\cr
{1\over \tau}= {1\over \tau_B}+{1\over \tau_A}, %\cr{}
\quad &\omega^2= {1-\gamma{Ap\over E}\over \tau_B+\tau_A},\cr{}
p(t)&= \int^t u(t') dt'
\end{align}
Although $\omega^2$ depends on $A,p$ and $E$, we can use an approximate
time dependent exponential ansatz $u\sim u_0 \exp[\lambda t]$. The
solutions to $\lambda$ are:
\[\lambda_\pm={-1\pm \sqrt{1- 4 \tau \omega^2}\over 2\tau}\]
When $\omega^2>0$ and $1-4 \tau \omega^2<0$ there will be oscillatory
solutions. For example when $\gamma{Ap\over E}<-1$, which only happens
for negative $\gamma$ we have such oscillatory solutions. This is
consistent with the simulations which showed the oscillatory behavior
was in the $\alpha\beta<0$ quadrants. For the stability, however we care
about the real solutions.

When $\omega^2<0$, which happens when $\gamma{Ap\over E}>1$, we will
have two real solutions with opposite signs. The presence of the
positive root signals an instability because the solution diverges. For
a delta function shock of magnitude $f$ at $t=0$ we found that:
\[E_0\to E_0(1+f)\]
Having initially scaled to $E_0=A_0=p_0=1$, the condition for existence of the positive root becomes:
\[t=0:\quad \gamma> {E\over Ap}= (1+f)\]
This dependence on the shock magnitude is normal, as a strong enough
kick can kick a particle out of a local minimum.  The shock can be
arbitrarily small and therefore the absolute condition for stability is
as we anticipated
\begin{equation}
\large \mbox{\bf unstable at:} \quad \gamma >1 \label{eq:unstable}
\end{equation}

Now the question is, which solution does the system pick when it is shocked. The return $\ro_t p$ is
\[\ro_t p(t) = u(t)= u_+ e^{\lambda_+ t}+u_- e^{\lambda_- t} \]
Since at $t=0$ the initial conditions dictated $\ro_t p(0)=0$ we have
\[u_+=-u_-\]
And therefore both solutions appear with equal strength. It follows that
whenever one of the solutions ($u_-$ in our case) is positive the
solution diverges. When $f>0$ a bubble forms and grows exponentially and
when $f<0$, because our variables are non-negative, the price just
crashes to zero.  This proves that the sufficient condition for
stability is $\gamma<1$. Also note that the nonlinear terms are all
proportional to $\ro_t p$ and therefore at $t=0$
\[O\pa{u^2(0),\ro_t u^2(0)}=0\]
and so the solution is exact at $t=0$.

\subsection{Validity of perturbation theory near the phase transition}

For the above solution to be valid we must confirm that the corrections
are small. We must find a small parameter that exists in the neglected
terms which allows perturbative solutions to be viable. We had two sets
of nonlinearities: (1) $O\pa{(\ro_t p)^2}$; (2) $\gamma Ap/E$.

\subsubsection{the non-linearity $O\pa{(\ro_t p)^2}$}
First let
us examine the nonlinear terms in $O\pa{(\ro_t p)^2}$.  Note that the
instability happens when the larger root $\lambda_-$ becomes
positive. Thus near the transition we have
\begin{align}
4\tau\omega^2&\ll1 \cr
\lambda_+&\approx -{1\over \tau} + \omega^2 \cr{}
\lambda_-&\approx - \omega^2
\end{align}
And so being close to the phase transition means $\lambda_-\ll
1/\tau$. The consequence of this is that for $O\pa{(\ro_t p)^2}$ we get
(using the $u_+=-u_-$ found above)
\begin{align}
\tau_A\ro_t)u=&\tau_A u_+\pa{\lambda_+ e^{\lambda_+ t}-\lambda_- e^{\lambda_- t}} \cr{}
\approx & \tau_A u_+\pa{\lambda_+ e^{\lambda_+ t}-\lambda_- e^{\lambda_- t}} \cr{}
O\pa{u^2}=&\tau_B {u(1+\tau_A\ro_t)u \over p}\cr &-\alpha \tau_B
{\pa{(1+\tau_A\ro_t)u}^2\over p}\cr
\approx & \tau_B {u(1+\tau_A u_+(\lambda_+ -\lambda_-) )u \over p}\cr
&-\alpha \tau_B {\pa{(1+\tau_A\ro_t)u}^2\over p}\cr
\end{align}

\subsubsection{The non-linearity $\gamma Ap/E$ }
We wish to examine if the assumption that in the stable regime $\ro_t A, \ro_t p, \ro_E$ remain small is a consistent assumption, thus making perturbative expansion valid. Any term above non-linear in $\ro_t A, \ro_t p, \ro_E$ is thus higher order in this approximation. We wish to find the part of $\gamma Ap/E \ro_t p$ that is linear in the first time derivative. In the stable regime changes are slow and thus a short time after the shock we can expand the variables in Taylor series near $t=0$. Again, we will rescale the variables at $t=0$ to $E_0=p_0=A_0=1$. Using the \eqref{eq:ddE} $\ro_t E= A\ro_t p$  we get
\begin{align}
{A(t)p(t)\over E(t)}\ro_t p & = {A_0p_0+ t (\ro_t A_0 p_0 + A_0 \ro_t p_0)\over E_0 + t A_0\ro_t p_0} \ro_t p \cr
&\approx {1\over E_0}\pr{A_0p_0 + t (\ro_t A_0 p_0 )} \ro_t p \cr
&= \ro_t p + O(\ro_t A_0 \ro_t p) \approx \ro_t p
\end{align}

Thus the assumption of smallness of the derivatives is consistent and we may use perturbation theory and safely discard the non-linear terms in finding the stability conditions. This way the stability condition is just having a positive $\omega^2$ in \eqref{eq:unstable}.
One can also check the stability by explicitly using the exponential ansatz found above as is given in what follows.

%   }%%%%%%%%%%%%%%%%
%%%%%%%%%%%%%% OLD derivations

%\outNim{ 120614

\section{Proof for $\gamma=1$ using properties of the phase transition\label{ap:exp}}

\outNim{%%%%%%%%%%%%%%%%%%%%%%%%%
%%%%%%%%%%%%%%%%%%%
%%%%%%%%%%%%%% OLD DERIVATIONS

Plugging these into the equations \eqref{eq:ddA}-\eqref{eq:ddE} we get
($\tau_A=\tau_B=T$)
\begin{align}
Tw_A^2+w_A-\beta w_E &=0\cr
Tw_p^2+w_p-\alpha w_A &=0\cr
w_E\ln w_E- (w_p+w_A)\ln w_p&=0\cr
w_E(1+f)-w_p&=0
\label{eq:west}
\end{align}

Now, before going into the details of the solutions, we will give a
simple argument about why the phase transition should happen at
$\alpha\beta=1$. Because we have no discontinuity that can appear on our
equations, we expect the phase transition to be smooth and similar to a
second order transition (consistent with our observations from the GIIPS
system). This implies that near the phase transition the effective
potential is changing from having a stable minimum to having an unstable
maximum. It is like a second order potential is switching signs. The
result of this smooth transition from concave to convex is that at the
phase transition the potential becomes very flat, before switching to
being convex. For this reason in second-order phase transitions the
forces, which are the gradients of the potential, become very small and
as a result the relaxation time for the system to reach its final state
becomes very large (again, just as we observed in the phase diagram of
the GIIPS system).

We can use this fact about relaxation times being large to assume that
the acceleration terms in the first two equation of \eqref{eq:west} must
be small. In other words if we write $T_{A,p,E}\equiv 1/ w_{A,p,E}$
then \[T_{A,p,E}\ll T \] which is another way of saying the relaxation
time is large. Let's define the dimension-less rescaled exponents
\[\eps_{A,p,E}\equiv T w_{A,p,E}={T\over T_{A,p,E}},\]
and the condition for the phase transition curve becomes
\begin{equation}
\eps_{A,p,E}\ll1. \label{eq:wll1}
\end{equation}

Multiplying the first two lines of \eqref{eq:west} by $T$ results in
\begin{align}
\eps_A^2 + \eps_A -\beta \eps_E \approx  \eps_A -\beta \eps_E=0\cr
\eps_p^2 + \eps_p -\beta \eps_A \approx  \eps_p -\alpha \eps_A=0
\end{align}
Using the last equation of \eqref{eq:west} we have
$\eps_E=\eps_p/(1+f)$. Thus combining these two, the condition for phase
transition becomes:
\begin{align}
\eps_p =\alpha \eps_A &= {\alpha \beta \over 1+f} \eps_p\cr
\Rightarrow \gamma=\alpha\beta  &=(1+f)
\end{align}
This is the curve of the phase transitions for shock magnitude
$f$. Taking the external shock to zero $f\to0$ yields the position of
the phase transition
\[\mbox{Phase transition at:}\quad \gamma =1\]
Note that this argument also holds if $\tau_A\ne \tau_B$. At the phase
transition the relaxation time is larger than all time-scales involved so
$T_{A,P,E}\gg \tau_{A,B}$ and thus again the $w^2_{A,p,E}$ terms become
negligible and we get
\begin{align}
 w_A -\beta w_E=0\cr
w_p -\alpha w_A=0,
\end{align}
and again the same result is obtained by plugging in $w_E=w_p/(1+f)$.

}%%%%%%%%%%%%%%%%%%%%%%%%%%%%%
%%%%%%%%%%%%%%%%%%%%%%%%%%%%%%

Since we have coupled second order equations, the solutions may be
estimated using an exponential ansatz as follows.
Equations \eqref{eq:ddA} and \eqref{eq:ddp} are second order and therefore will naturally have two solutions for $A$ and $p$. Also, since $\ro_t E= A\ro_t p$, $E$ will also have two modes. Therefore the exponential ansatz must have at least two exponents.
Thus for each of the three variables $X=E,p,A$ we have:
\[X\sim X_0 +X_1\exp[w_{X1} t]+X_2\exp[w_{X2} t]\]
%\[E\sim E_0 \exp[w_E t]+E_1,\quad A\sim A_0 \exp[w_A t]+A_1,\quad  p\sim p_0 \exp[w_p t]+p_1\]
%
%\newpage

In principle the exponents can be time-dependent, but we will first
try and see f there are asymptotically exponential solutions. Thus we
assume that they vary slowly with time. By choosing the units of $E,p,A$ to be such that at $t=0+\epsilon$, $p=A=1$ and the shocked equity is $E=1+f$,
the boundary conditions that we had become:
\[%X_0+X_1+X_2=0,
A_0=1-A_1-A_2,\quad p_0=1-p_1-p_2,\quad E_0=1+f-E_1-E_2\]
and:
\begin{align}
\ro_t E(0)&=0 =w_{E1}E_1+w_{E_2}E_2=0\cr
&= A\ro_t p= w_{p1}p_1+w_{p_2}p_2 \cr
\ro_t A(0)&= {\beta\over \tau_B} \ln (1+f)= w_{A1}A_1+w_{A_2}A_2
\end{align}
At the phase transition we expect the greater exponents, which we take to be $w_{X2}$, to become small relative to other time-scales in the problem, i.e. $\tau_A,\tau_B$, and change sign from negative (which would result in exponential decay) to positive (which results in divergence of $E,p,A$). This means that close to the phase transition:
\[|w_{X2}|\ll |w_{X1}|, \quad w_{X2}\ll {1\over \tau_A}+{1\over \tau_B}\]
From the initial conditions, this results in:
\begin{align}
|E_1|&= |{w_{E2}\over w_{E1}}E_2| \ll |E_2|,  \cr
\Rightarrow E_0&=1+f-\pa{1-{w_{E2}\over w_{E1}} }E_2\approx 1+f-E_2\cr
|p_1|&\ll |p_2| ,\qquad  \Rightarrow p_0 \approx 1-p_2
\end{align}

For $A$ we have a little more details.
\[A_1= {\beta\over w_{A1}\tau_B} \ln (1+f)- {w_{A2}\over w_{A1}}A_2 \approx  {\beta\over w_{A1}\tau_B} \ln (1+f) %\approx \beta f
\]
Which for small shocks $f\ll1$ reduces to:
\[A_1 \approx  {\beta\over w_{A1}\tau_B}f\]
Now back to the equations \eqref{eq:ddA}-\eqref{eq:ddE}.
First let us reexamine the
third equation \eqref{eq:ddE}. The effect of a delta function shock
$f(t)=f_0\delta(t)$ is the above $\ro_t A$ and $E(+\epsilon)= (1+f_0)$. Since $|w_{X2}|\ll |w_{X1}$ and $w_{X1}<0$ we can neglect $\exp[w_{X1}t]$. The last equation becomes:
\begin{align}
\ro_tE&= w_{E2}E_2\pa{e^{w_{E2}t}-e^{w_{E1}t}}\approx w_{E2}E_2e^{w_{E2}t}\cr
&= A\ro_t p = \Big(1+A_1\pa{e^{w_{A1} t}-1}\cr
&+A_2\pa{e^{w_{A2} t}-1}\Big) w_{p2}p_2\pa{ e^{w_{p2} t}- e^{w_{p1} t}}\cr
&\approx \pa{1+{\beta\over w_{A1}\tau_B}f +A_2\pa{w_{A2} t}} w_{p2}p_2 e^{w_{p2} t}\cr{}
\end{align}
For arbitrary $t$ this relation can only hold if $w_{E2}=w_{p2}$. Thus we define:
\[w\equiv w_{E2}=w_{p2}\]

Let us also get an estimate for $w_{A1}$, the smaller exponent in $A$. We will go very close to the transition line where $w_{X2}\approx 0$. From Eq. \eqref{eq:ddA} we have:
\[{(\tau_B \ro_t +1)\ro_t A\over A}=\beta {\ro_t E\over E}\approx \beta {w E_2 e^{wt}\over E} \approx 0 \]
With an exponential ansatz the left hand side is:
\[(\tau_B w_A +1)w_A =0\]
The greater root is $w_{A2}=0$ and the smaller root is $w_{A1}=1/\tau_B$. Even away from the transition line we approximately have:
\[w_{A1}+w_{A2}\approx {1\over \tau_B}\]
Thus we can approximate the expression for $A_1$ to:
\[A_1\approx {\beta\over \tau_B} \ln (1+f)\approx \beta f\]
This way Eq. \eqref{eq:ddE} becomes
\begin{align}
E_2&\approx \pa{1+\beta f +A_2\pa{w_{A2} t}} p_2 \approx (1+\beta f) p_2
\end{align}
\outNim{ %%%%%%%%
\begin{equation}
w_{E2}(1+f)=w_{p2} \label{eq:wewp}
\end{equation}
} %%%%%%%%
Eq. \eqref{eq:ddA} becomes
\begin{align}
{(\tau_B w_{A2}+1)w_{A2}A_2 e^{w_{A2}t} \over 1+\beta f \pa{e^{t/\tau_B}-1}+A_2 w_{A2} t}&\approx  \beta {wE_2 e^{wt}\over 1+f + E_2 w t}
\end{align}
Which again only holds if $w_{A2}=w$. Thus
\[w_{A2}=w_{E2}=w_{p2}=w\]
Again, note that the condition for being close to the transition point was:
\[w\ll {1\over \tau_A}+{1\over \tau_B}\]
Discarding higher than linear order terms in $w$ and looking at times $t/\tau_B\gg 1$ yields:
\begin{align}
{ A_2 \over 1-\beta f }&=  \beta { E_2 \over 1+f }+O(w)
\end{align}
Performing the same procedure on Eq. \eqref{eq:ddp} results in (since $w\ll {1\over \tau_{A,B}}$)
\begin{align}
&{(\tau_A w+1)wp_2 e^{wt}\over 1 + p_2 w t}\approx  \alpha { wA_2 e^{w t} \over 1+\beta f \pa{e^{t/\tau_B}-1}+A_2 w_{A2} t}  \cr
p_2&=  \alpha { A_2  \over 1-\beta f }+O(w)= \alpha \beta {E_2\over (1+f)}+O(w)\cr{}
&\approx \gamma {1+\beta f\over 1+f} p_2 +O(w)
\end{align}
And so, the condition for the phase transition becomes:
\[\gamma = {1+f\over 1-\beta f}\]
%
%\newpage
%
Now, taking the shock to zero $f\to 0$ results in a phase transition at:
\begin{align}
\mbox{Phase~Transition at:~}\gamma = 1
\end{align}

%120614 }

%120714}

% \bibliographystyle{plain}
% \bibliography{references}

\end{document}